# Controlling Intensity and Phase of Terahertz Radiation with an Optically Thin Liquid Crystal-Loaded Metamaterial

O. Buchnev[1], J. Wallauer[2], M. Walther[2], M. Kaczmarek[3], N. I. Zheludev[1,4] and V. A. Fedotov[1,*]

[1] *Optoelectronics Research Centre and Centre for Photonic Metamaterials, University of Southampton, SO17 1BJ, UK*
[2] *Department of Molecular and Optical Physics, University of Freiburg, D-79104, Germany*
[3] *Department of Physics and Astronomy, University of Southampton, SO17 1BJ, UK*
[4] *Centre for Disruptive Photonic Technologies, Nanyang Technological University, 637371, Singapore*

[*] e-mail: vaf@orc.soton.ac.uk

**Abstract:** We experimentally demonstrate intensity and phase modulation of terahertz radiation using actively controlled large-area planar metamaterial (metafilm) hybridized with a 12 μm thick layer of a liquid crystal. Active control was introduced through in-plane electrical switching of the liquid crystal, which enabled us to achieve a reversible single-pass absolute transmission change of 20 % and a phase change of 40 deg at only 20 V.

Efficient active control of terahertz radiation is one of the main challenges of the terahertz technology. The mainstream solutions that have been demonstrated during the past decade include the use of semiconducting structures and liquid crystals (LC) [1,2,3,4,5]. Despite the proven advantage of being broadband these solutions have suffered from a number of drawbacks. In particular, superconductor-based THz modulators have a very small active area and are able to modulate transmission by only a few percent [2] and normally require cryogenic temperatures [1]. Whereas LC optical cells are unable to control the intensity of terahertz radiation, cannot be thinner than several hundred microns (due to the relatively low THz birefringence of liquid crystals), and also require bulky magnets [3] or a driving voltage in excess of 100 V [4,5].

An intriguing way of improving the performance characteristics of such active THz devices has emerged recently with the advent of metamaterials, artificially structured electromagnetic materials that are designed to manipulate light in ways no natural materials can [6]. Metamaterials have advanced rapidly over the past few years and are now expected to have a major impact across the entire range of technologies where electromagnetic radiation is used, ranging from RF and microwave antennas to photonics and nanophotonics [7]. The metamaterial concept has not only brought to life such exotic optical effects as artificial magnetism [8,9], negative refraction [10] and cloaking [11], but also enabled the dramatic enhancement of light-matter interaction leading to amplified absorption [12], giant polarization rotation [13] and slow light propagation [14,15]. The enhanced light-matter interaction is a direct consequence of narrowband metamaterial resonances, which can be engineered for virtually any frequency, and it is this property of the metamaterials that has been used in the recent demonstrations of efficient electro-optical and all-optical room-temperature THz modulators incorporating active semiconducting components [16,17,18,19,20].

In this Letter we show experimentally that electrically tunable birefringence of an optically thin layer of a liquid crystal, which is too weak to produce any noticeable transmission effect alone, can yield a very efficient radiation control mechanism when combined with the strong resonant response of a THz planar metamaterial (metafilm). The demonstrated active LC-loaded metamaterial hybrid enables the control of both intensity and phase of the transmitted terahertz radiation and requires only a moderate driving voltage for its operation.

Our metafilm was based on the so-called fish-scale (FS) pattern [21], a regular array of continuous meandering metallic wires (see Fig. 1a). The pattern was etched using high-resolution photolithography on a 220 nm thick aluminum film that had been sputtered beforehand onto 500 μm thick oxidized silicon substrate (the thickness of silicon dioxide layer was 10 μm). The width of the resulting aluminum tracks was 5 μm. The fabricated metamaterial array had a square unit cell and a period of 100 μm, which made it non-diffracting below 1.5 THz for any angle of incidence. The overall size of the sample was 12 mm × 12 mm.

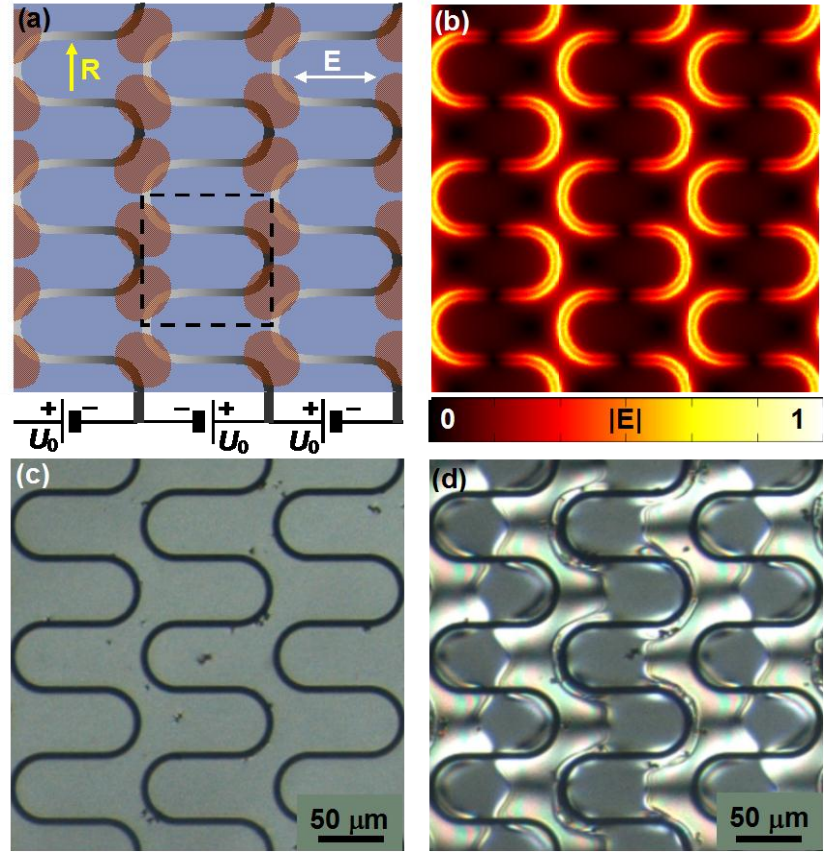

**Figure 1.** (Colour online) (a) Schematic of fish-scale planar metamaterial and electric circuitry providing the driving voltage $U_0$ for in-plane control of liquid crystal. Shaded areas indicate localizations of in-plane component of applied electrical field. (b) Calculated distribution of resonant electric field induced in metamaterial by normally incident plane wave at 0.9 THz. Panels (c) and (d) show images of liquid crystal-loaded metamaterial made under polarization optical microscope while applying 0 V and 20 V respectively. Bright domains correspond to in-plane switching of liquid crystal. For the purpose of imaging metamaterial was fabricated on a transparent substrate (quartz).

Transmission response of the metafilm was characterized at normal incidence in the 0.4 – 1.2 THz range of frequencies using the standard terahertz time-domain spectroscopy (THz-TDS) technique. The polarization of the incident wave was set parallel to the straight sections of the meanders, as illustrated in Fig. 1a. Despite the vanishing thickness of the aluminum pattern the metafilm exhibited a pronounced transmission stop-band centered at around 0.9 THz (see Fig. 2a). The stop-band corresponded to the first fundamental resonance of FS metamaterial when the total electrical length of the metallic track confined within the unit cell became equal to the wavelength of radiation at the structure's interface [21]. The resonance was accompanied by strong phase dispersion for the transmitted wave (see Fig. 2b). For what follows below it is also important to note that the electric field of the resonantly induced current mode was localized at the curved sections of the meanders, as shown in Fig. 1b.

To be able to load FS metamaterial with liquid crystal we placed a quartz cover slide 12 μm above the plane of the aluminum pattern, which rendered the resulting structure an optically thin cell. The cell was filled with highly birefringent nematic LC (1825 [22]) and sealed to prevent leakage of the liquid crystal. The surface of the metafilm and the inner surface of the cover slide were coated with a polymer (PI-2525 from HD MicroSystems) and rubbed in the direction orthogonal to the straight sections of the meanders, as illustrated in Fig. 1a. The latter promoted uniform alignment of LC molecules in the cell orthogonal to the incident polarization. The presence of ordered LC layer red-shifted the resonance by ~ 0.12 THz.

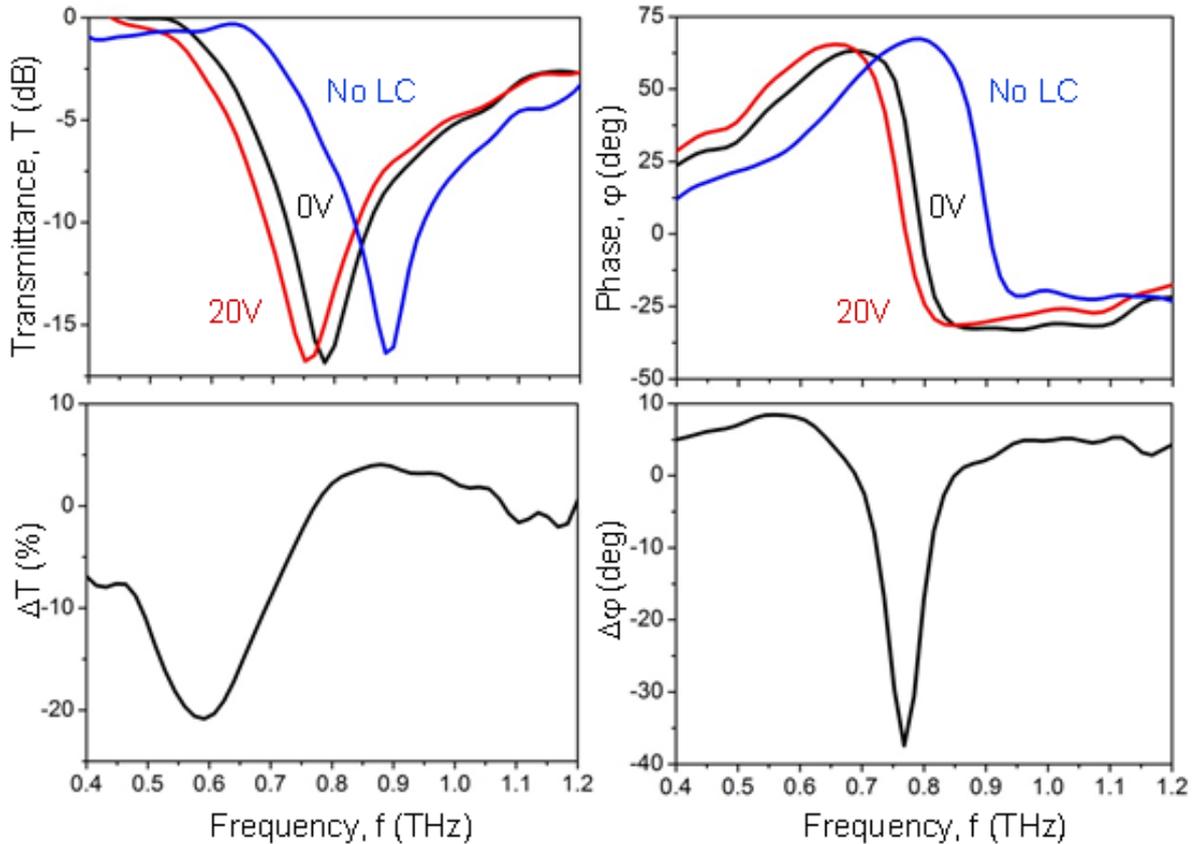

**Figure 2.** (Colour online) Panels (a) and (b) show transmission spectra and phase dispersions for both empty and LC-loaded metamaterials. Panels (c) and (d) show spectral dependences for absolute changes of transmission intensity and phase achieved at 20 V.

The role of the cell's control electrodes was played by the metallic network of the metafilm itself. The meandering strips were split into two groups and connected to an electric potential of opposite signs, as shown in Fig. 1a. The resulting configuration of the applied electric field would ensure re-orientation of LC molecules in the plane of the structure with the maximum effect occurring near the curved sections of the meanders due to the proximity of the electrodes (where indicated by shaded areas in Fig. 1a). That was also confirmed experimentally using polarization optical microscopy (see Figs. 1c – 1d). To avoid the appearance of double-charged layers we used the common driving scheme, where the control voltage was alternated in the form of square waves with a frequency of 1 kHz. Importantly, LC domains exhibiting the strongest molecular re-orientation (and therefore the largest local change of the refractive index) were seen to overlap with the areas of high E-field concentration produced by the resonant current mode since the latter were also confined to the curved sections of the meanders (as evident from Fig. 1b). This had ensured efficient tunability of the metamaterial resonance achieved in the presence of an in-plane electric field. In particular, by applying a maximum of 20 V between the meanders we were able to red-shift the resonance frequency by 4 %. The demonstrated tuning range is close to the absolute theoretical limit of 6 %, which was determined assuming that the LC layer could be fully switched to an isotropic state with the highest refractive index $n_e$.

The electrical tuning of the metamaterial resonance resulted in a change of the structure's overall transmission across the entire spectral domain, with a maximum absolute difference in the transmitted intensity $|\Delta T|_{max} = |T(20\text{ V}) - T(0\text{ V})| \approx 20$ % observed at around 0.60 THz (see Fig. 2c). The shift of the resonance frequency also produced an offset in the phase dispersion, introducing a substantial change in the retardation of the transmitted wave. The maximum absolute change corresponded to $|\Delta \varphi|_{max} = |\varphi(20\text{ V}) - \varphi(0\text{ V})| \approx 40$ deg and was achieved at 0.77 THz (see Fig. 2d). It is important to note at this point that to produce the same effect in a conventional LC optical cell the physical thickness of the latter would have to be $h = \Delta \varphi \, \lambda / 2\pi \, \text{Re}(n_e - n_o) = 230$ μm, which is almost 20 times larger than the thickness of the LC layer used in our hybrid metamaterial structure. By varying the amplitude of the driving voltage, $U_0$, in the range from 0 to 20 V we could control both the intensity and phase of the transmitted wave in a continuous and reversible fashion observing monotonous dependencies of $|\Delta T|$ and $|\Delta \varphi|$ on $U_0$, as evident from Fig. 3a.

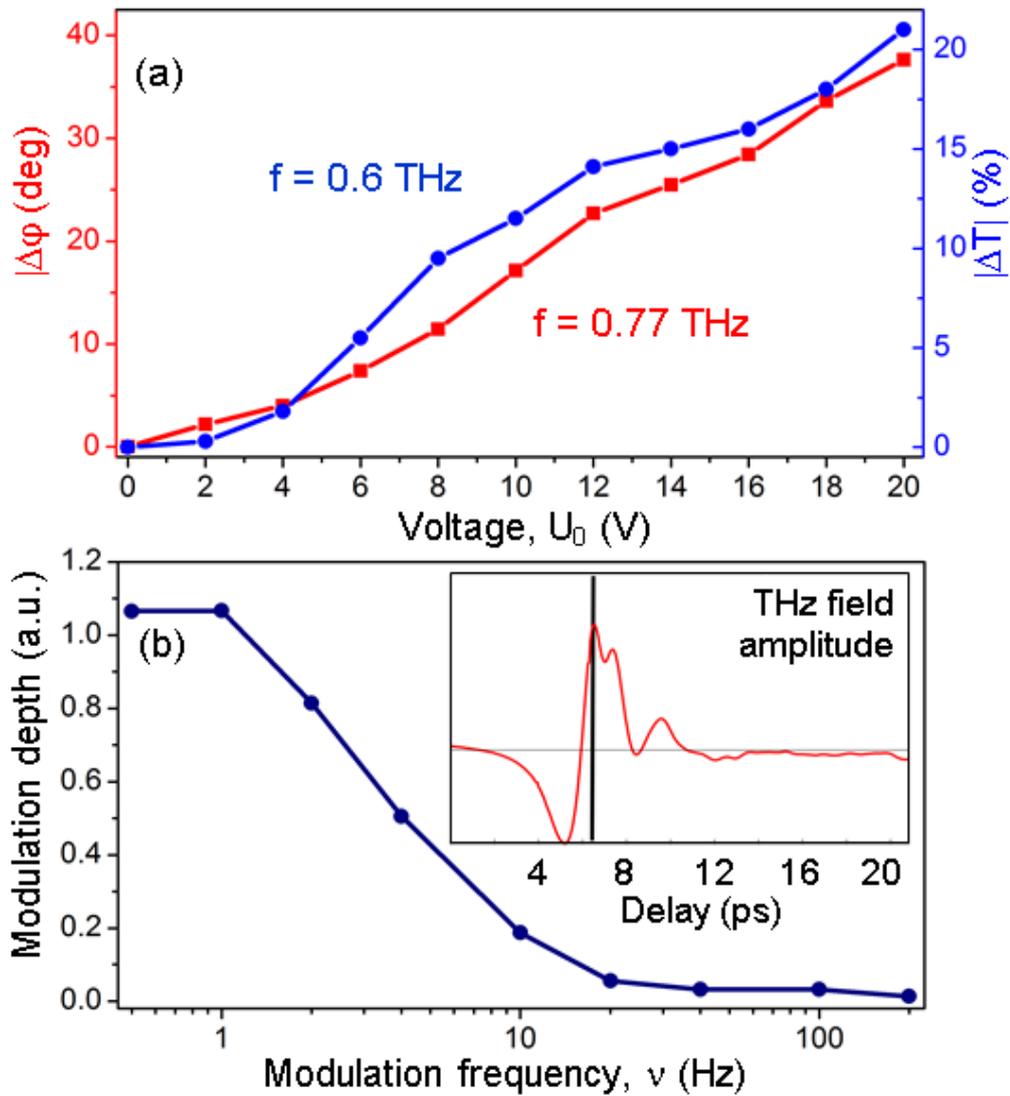

**Figure 3.** (Colour online) Maximum changes in transmission intensity and phase of LC-loaded metamaterial measured as functions of applied voltage (a). Depth of amplitude modulation of TDS THz pulses measured for different modulation frequencies (b). Inset show temporal profile of TDS THz pulse

The response dynamics of the hybrid metamaterial was determined by the switching and relaxation times of its liquid-crystal sub-system. While for most of the liquid crystal-based devices the switching time is typically of the order of few tens of milliseconds (unless a special driving scheme is used), the relaxation time is usually substantially longer and depends on the size and shape of the liquid crystal volume, spatial deformation of the LC director, its boundary alignment layers etc. To determine the relaxation time in our hybrid system we studied its transmission response in the time domain. The response was modulated by chopping the driving voltage with 50 % duty cycle and frequencies ranging from 0.5 to 200 Hz. The resulting transmission modulation was characterized as the voltage-induced change of the amplitude of TDS THz pulses exiting the metafilm. Figure 3b shows modulation depth for the amplitude of THz pulses measured as a function of the modulation frequency. One can see that below 1 Hz the transmission modulation is at a maximum and remains constant, but at higher frequencies it gradually becomes weaker and completely vanishes above 20 Hz. Based on this dependence, which has a roll-off at around $\nu_{off} = 4$ Hz, we estimated the relaxation time for the hybrid metamaterial structure $\Delta\tau_r$ to be around 125 ms ($\Delta\tau_r \approx 1/2\nu_{off}$). Surprisingly, its response appears to be faster by almost a factor of 2 than that of a commercial liquid crystal optical cell of the same thickness [23]. We attribute this to the difference in the nature and implementation of the switching modes used in both cases, which in our case corresponded to the in-plane rather than volume switching engaging only a fraction of the bulk of the liquid crystal (as shown in Fig. 1d).

In summary, we experimentally demonstrated efficient intensity and phase modulation of terahertz radiation using an actively controlled metafilm combined with a liquid crystal layer only 12 μm thick. The absolute change in intensity and phase achieved for a single pass transmission were 20 % and 40 deg respectively. The demonstrated LC-metamaterial hybrid for the first time exploited the in-plane LC switching mode, which substantially simplified the design of the structure and allowed it to operate in the transmission regime, as well as enabled the reduction of the driving voltage down to a few tens of volts. That set our approach apart from recently proposed schemes for active control of THz metamaterials with liquid crystals [24, 25].